\documentclass[letterpaper, 10 pt, conference]{ieeeconf}  % Comment this line out
\IEEEoverridecommandlockouts                              % This command is only
\usepackage{amsmath} % assumes amsmath package installed
\usepackage[linesnumbered,ruled]{algorithm2e}
\usepackage{amssymb}  % assumes amsmath package installed
\usepackage{graphicx}
\usepackage{tabularx,booktabs,hyperref}
\newcolumntype{C}{>{\centering\arraybackslash}X} % centered version of "X" type
\title{\LARGE \bf
Optimizing Market Making using Multi-Agent Reinforcement Learning
}

\author{ \parbox{3 in}{\centering Yagna Patel\\
        {\tt\small yagna.patel@berkeley.edu}}
        \parbox{3 in}{ 
        }
}

\begin{document}

\maketitle
\thispagestyle{plain}
\pagestyle{plain}

\begin{abstract}
In this paper, reinforcement learning is applied to the problem of optimizing market making. A multi-agent reinforcement learning framework is used to optimally place limit orders that lead to successful trades. The framework consists of two agents. The macro-agent optimizes on making the decision to buy, sell, or hold an asset. The micro-agent optimizes on placing limit orders within the limit order book. For the context of this paper, the proposed framework is applied and studied on the Bitcoin cryptocurrency market. The goal of this paper is to show that reinforcement learning is a viable strategy that can be applied to complex problems (with complex environments) such as market making.

\end{abstract}

\section{Introduction}
Algorithmic trading, and in particular high-frequency algorithmic trading (HFT), has gained immense popularity in the recent decade. With advances in hardware and software, algorithmic trading has rapidly become the norm. The increasing popularity of machine learning has slowly made its way to financial markets \cite{c1}, where it is primarily used to predict price movements of assets. However, there are a number of challenges that these classic machine learning techniques entail:
\begin{enumerate}
    \item \textit{Prediction time: }In machine learning, model complexity can have an impact on prediction time. Many times, in the supervised learning setting, neural networks are used to make predictions. Due to the computational complexity that comes with these models, as the model complexity increases, the decision time also increases \cite{c2}. In the HFT setting, by the time the model makes a prediction, it may already be too late to take the predicted action. The problem then becomes, how can these added latency costs be incorporated into our prediction? 
    \item \textit{Prediction accuracy: } Financial markets are second-order chaotic systems, i.e. they are highly unpredictable since markets can \textit{respond} to predictions. The general rule of thumb in finance is that the historical performance of an asset does not predict the future performance of the asset, i.e. forecasting and predicting the market from historical performance alone is virtually impossible. In fact, markets are sometimes compared to random walk processes \cite{c10}. Therefore, approaches that solely rely on the historical performance of an asset are likely to have low prediction accuracy.
    \item \textit{Policy optimization: }Suppose that our model predicted a $55\%$ chance of increase and a $45\%$ chance of decrease for an asset. How can our model's prediction be converted into an action? An optimized policy and certain decision thresholds are needed to turn the prediction into an action. However, coming up with such a policy is usually done by hand. The development and optimization of these policies are commonly done by mix of humans and computers. Therefore, if the market suddenly shifted, these policies would likely not be able to adapt.
\end{enumerate}
\section{Related Work}
The concept of reinforcement learning is relatively new in finance. There has been little research done into whether or not reinforcement learning is a viable approach for market making. The optimization problem of market making is a complex problem \cite{c11}, and reinforcement learning is not a common approach used to solve it. Multi-agent approaches to stock trading have been taken previously \cite{c9}. However, they do not account for placing limit orders. The general concept of the micro-agent (see section 3) has been shown by Nevmyvaka et al. \cite{c3}, and further extended by Juchli \cite{c4}. These papers present the problem of optimally buying (or selling) an asset over a fixed time horizon. This is similar to the idea of the micro-agent. The work for the micro-agent is substantially based, and builds upon the ideas presented in these two papers. 
\section{Problem Formulation}
In this paper, a less common approach of reinforcement learning is utilized to create an optimal market maker. More specifically, a multi-agent approach (with two agents) is used:
\begin{enumerate}
    \item \textit{Macro-agent: }The macro-agent is given minute tick data (data at a macro-level) and makes the decision to buy, sell, or hold the asset. 
    \item \textit{Micro-agent: }The micro-agent is given order book data (data at a micro-level) and makes the decision of where to place the order within the limit order book.
\end{enumerate}
\begin{figure}[!ht]
    \centering
    \includegraphics[scale=0.15]{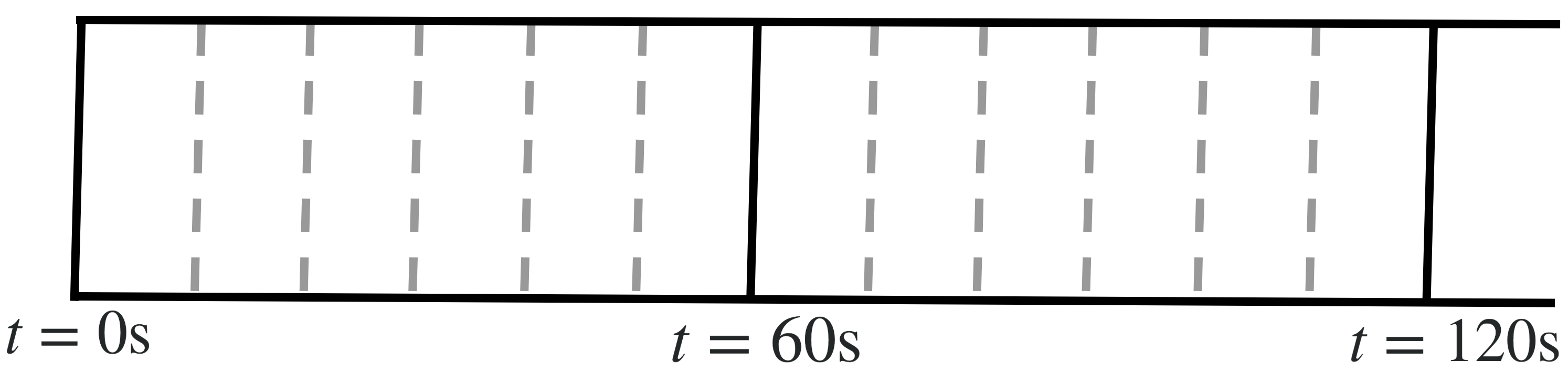}
    \caption{Time-steps where agents take actions. The Macro-agent takes actions at every bold black line, while the Micro-agent takes actions between the dashed lines.}
    \label{fig:execution}
\end{figure}
At the start of every minute, the macro-agent makes the decision to buy, sell, or hold the asset. A running count of how many buys it chooses to make is kept. The total number of buys before the sell is fed into the micro-agent, i.e. all collected assets are exhausted on each decision to sell. Additionally, on each decision to sell, the count is reset. Within the $60$ second time horizon that follows, the micro-agent attempts to exhaust its held assets by only placing limit orders. The agent may place multiple limit orders, however, the it is limited to a single order placement every $10$ seconds, as depicted in Fig. \ref{fig:execution}.

\subsection{Data Collection}
\begin{figure}[!ht]
    \centering
    \includegraphics[scale=0.30]{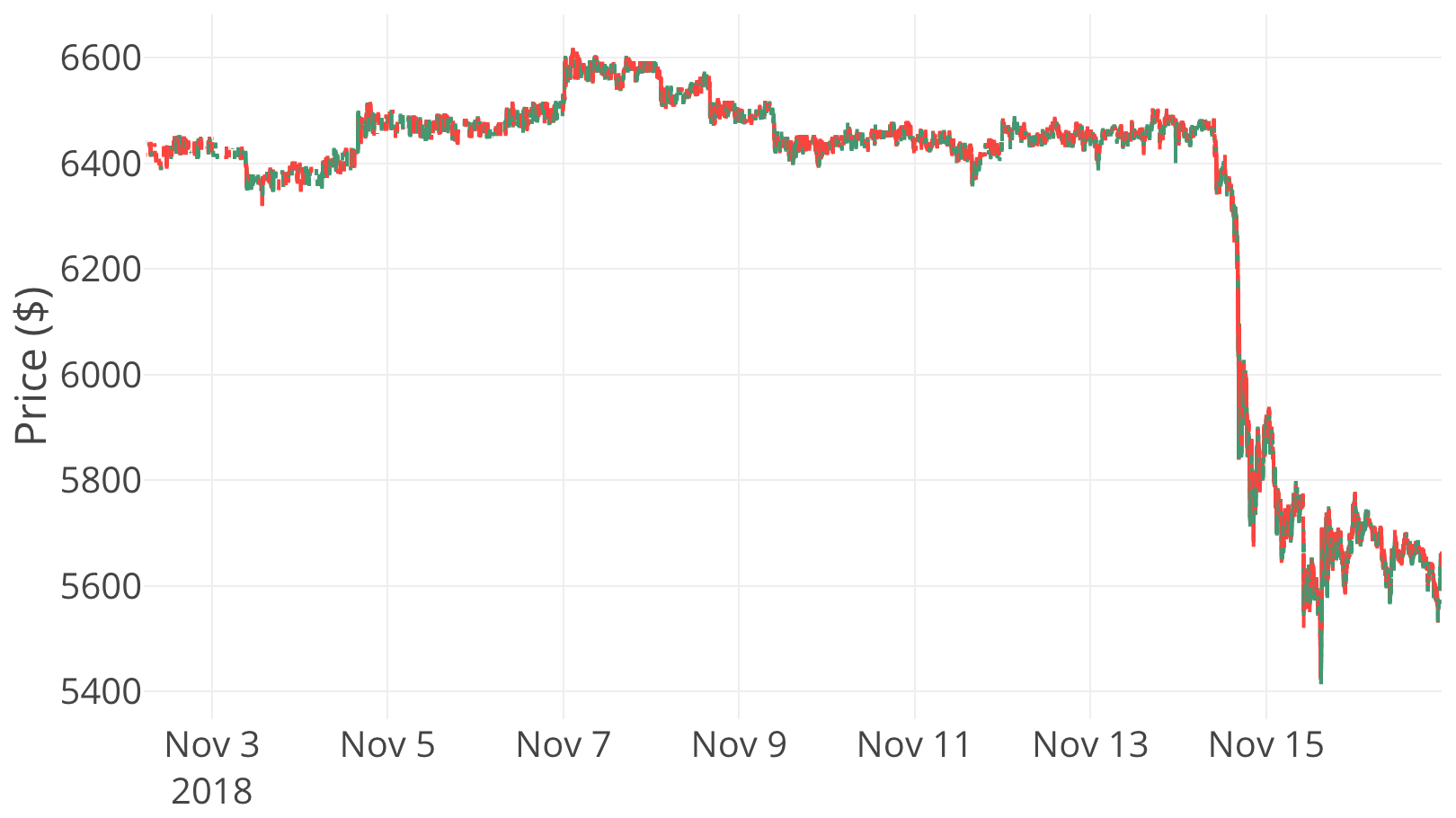}
    \caption{Bitcoin Minute tick data collected over 2018-11-02 to 2018-11-17}
    \label{fig:btc_min_large}
\end{figure}

Historical order book and minute tick data for markets is not readily available. The first step is to collect this data. The Bitcoin cryptocurrency market is used in this study for the reason that its data is readily available. All trade, bid, and ask data is collected by subscribing to Bittrex's WebSocket in the date range of November 2nd, 2018 to November 17th, 2018. With data provided by the WebSocket, a historical order book is constructed, i.e. a sequence of historical order book states over the time period. A minute tick dataset is then created using the trade data collected. In total, $41830629$ trade, bid, and ask data points, and $10945$ minute tick data points (Fig. \ref{fig:btc_min_large}) were collected for this study. However, the entire dataset was not used. Instead, only the most recent few days were chosen.

\section{Background: Brief Overview of Reinforcement Learning}
Reinforcement learning (RL) is a learning technique within sequential decision making where an agent learns to take actions optimally in an environment. Unlike supervised learning, in RL, the agent learns to take actions that maximize the reward it receives from the environment. At each time-step, the agent observes its state and takes an action based on the observation. The environment provides feedback on how well the action performed in the form of a reward. Many times these rewards can be delayed. For example, the decision to hold an asset may not yield an instant reward.

The process of an agent interacting with the environment is formalized as a Markov Decision Process (MDP). MDPs are used to describe the environment in the context of an RL problem. The importance behind MDPs in this problem is the Markov property that MDPs assume: the effects of some action taken in some state depends only on that state and not on prior states encountered. Market prices are Markovian in nature, i.e. the probability distribution of the price of an asset depends only on the current price and not on the prior history. Hence, it makes sense to formulate this problem as an MDP problem with the goal being to solve this MDP by finding an optimal policy. For each agent, the tuple $(\mathcal{S},\mathcal{A},P,r,\gamma)$ describes the problem:
\begin{itemize}
    \item $\mathcal{S}$ is the set of states.
    \item $\mathcal{A}$ is the set of actions.
    \item $P:\mathcal{S}\times \mathcal{A}\times \mathcal{S}\to\mathbb{R}$ is the transition probability distribution which determines how the environment changes based on the state and actions taken by the agent.
    \item $r:\mathcal{S}\to\mathbb{R}$ is the reward function.
    \item $\gamma\in(0,1)$ is the discount factor which determines the importance of future rewards. 
\end{itemize}
Let $R_t$ denote the sum of future rewards, i.e. $$R_t = r_{t+1}+r_{t+2}+\cdots+r_T$$ Then, at time-step $t$, the future discounted reward is then given by \begin{align*}R'_t&=R_{t+1}+\gamma\cdot R_{t+2}+\gamma^2\cdot R_{t+3}+\ldots+\gamma^T\cdot R_{T}\\&=\sum_{i=t}^T\gamma^{i}\cdot R_{t+i+1}\tag{4.1}\end{align*} where $T$ is the last time-step, potentially infinity.

States in the context of market making are much more complex than typical RL problems. In fact, the states encountered are most often not the complete states since there exist many other traders in the environment. Thus, the market making problem is a Partially Observable Markov Decision Process (POMDP). The state the agent observes, $\mathcal{S}'$, is some derivation of the true state, $\mathcal{S}$, i.e. $\mathcal{S}'\sim\mathcal{O}(\mathcal{S})$. Nevertheless, given a proper simulation of the environment, the agent should be able to optimize well against the unknowns in the environment.

\begin{figure}[!ht]
    \centering
    \includegraphics[scale=0.13]{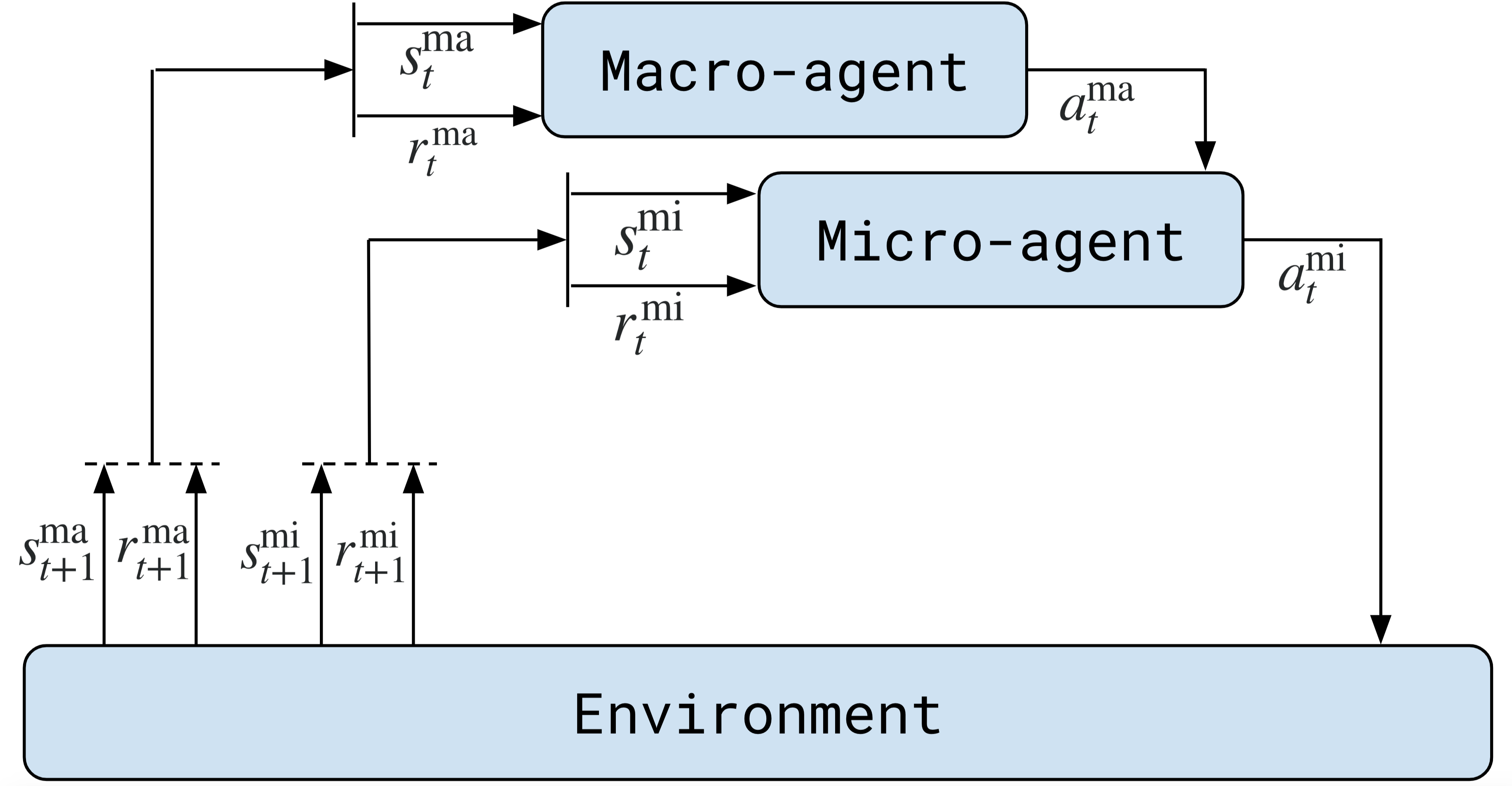}
    \caption{Multi-agent Reinforcement Learning Framework}
    \label{fig:marl}
\end{figure}

As previously mentioned, an RL agent continuously goes through a cycle of states by interacting with the environment. For this problem, the interaction between the agents and the environment is depicted in Fig. \ref{fig:marl}.

Both agents encounter states and output actions. The action determined by the macro-agent is fed into the micro-agent, which collectively turns into a single action: placing an order in the limit order book. The environment takes this action and outputs a reward and a next state. As noted in the problem statement, discrete time-steps are chosen (rather than continuous time-steps) for the reason that continuous time-steps would not be possible in the real world since the WebSocket data itself arrives at discrete time-steps. 

% \vspace{5pt}
% \noindent\textit{\underline{$Q$-Learning}}
% \vspace{2pt}
\subsection{$Q$-Learning}

The specific reinforcement learning algorithm used for both agents is deep $Q$-learning. Recall that the goal is to find an optimal policy $\pi:\mathcal{S}\to \mathcal{A}$, i.e. a function mapping between states and actions such that it maximizes the expected reward. $Q$-learning is a model-free value-based approach, which uses the action-value function $Q(s,a)$, where $s\in\mathcal{S}$ and $a\in\mathcal{A}$, to estimate the expected reward given some state-action pair. In other words, under policy $\pi$, the true value of being in state $s$ and performing action $a$ is given by \begin{align*}Q^\pi(s,a) = \mathbb{E}_\pi\left[R_t\mid \mathcal{S}_t=s,\mathcal{A}_t=a,\pi\right]\tag{4.2}\end{align*} Typically the $Q$-values are updated using the bellman equation, which are derived by expanding $(4.2)$:
\begin{align*}
    Q^\pi(s,a) &= \mathbb{E}_\pi\left[R_t\mid \mathcal{S}_t=s,\mathcal{A}_t=a,\pi\right] \\&= \mathbb{E}_\pi\left[R_{t+1}+\gamma R'_{t+1}\mid\mathcal{S}_t=s,\mathcal{A}_t=a\right] \\&= \sum_{s'}\sum_{r}\mathbb{P}(s',r\mid s,a)\left[r+\gamma\mathbb{E}_\pi\left[R'_{t+1}\mid\mathcal{S}_{t+1}=s'\right]\right] \\&= \sum_{s',r}\mathbb{P}(s',r\mid s,a)\left[r+\gamma Q^\pi(s',\pi_t(s'))\right]
\end{align*}
Based on this expansion, the update rule is given by: $$Q_{t+1}(s_t,a_t) = \sum_{s',r}\mathbb{P}(s',r\mid s,a)\left[r+\gamma\cdot Q^\pi_t(s',\pi_t(s'))\right]$$
Although this update rule would lead to an optimal policy, it is unfeasible for this problem since $\mathbb{P}(s',r\mid s,a)$ is unknown, i.e. the data collected does not provide a complete picture of the market environment. Instead we use the following update rule:
\begin{align*}Q_{t+1}&(s_t,a_t) \\&= (1-\alpha)\cdot Q_t(s_t,a_t) + \alpha\left[r_{t+1}+\gamma \max_a Q_t(s_{t+1},a)\right]\end{align*}
where $\alpha$ is a learning rate.

In this problem, a neural network is used to approximate the $Q$-value function where the input is the state (instead of state-action pairs), and the output are the $Q$-values for each of the actions. Thus, the $Q$-value function is parameterized by weights $\theta$, i.e. we assume that $$Q(s,a; \theta)\approx Q^*(s,a)$$

% \vspace{5pt}
% \noindent\textit{\underline{Exploration-Exploitation Trade-off}}
% \vspace{2pt}
\subsection{Exploration-Exploitation Trade-off}

In reinforcement learning, there is a trade-off between exploration and exploitation: how often do we want our agent to continue taking actions with the policy it has determined versus how often do we want our agent to explore taking new actions. To get a good balance between both \textit{worlds}, a decaying $\epsilon$-greedy approach is used. In this approach, the agent takes random actions with probability $\epsilon$ and takes the policy action with probability $1-\epsilon$, while $\epsilon$ decays over time. This strategy encourages the agent to explore more at the beginning, and exploit more at the end, i.e. stick to the policy it has computed.

% \vspace{5pt}
% \noindent\textit{\underline{Experience Replay}}
% \vspace{2pt}
\subsection{Experience Replay}

Another concept that is used is experience replay. In online learning, we receive data sequentially. Due to the nature of markets, there is always a concern for market shifts. If the market suddenly shifted, the agent may forget its past experiences as it aims to get a better advantage in the new states. By storing the agent's experiences in a memory buffer, and having the agent relive randomly sampled batches of experiences often, the high temporal correlation in the data is reduced. Additionally, training on these randomly sampled batches gives the \textit{feel} of training on iid data. Therefore, as in supervised learning problems, this would yield a better convergence for the function approximator to the optimal policy. Furthermore, this allows to update the parameters of the function approximator with a stochastic gradient descent approach.

\section{Details of the Macro-agent}
\begin{figure}[!ht]
    \centering
    \includegraphics[scale=0.30]{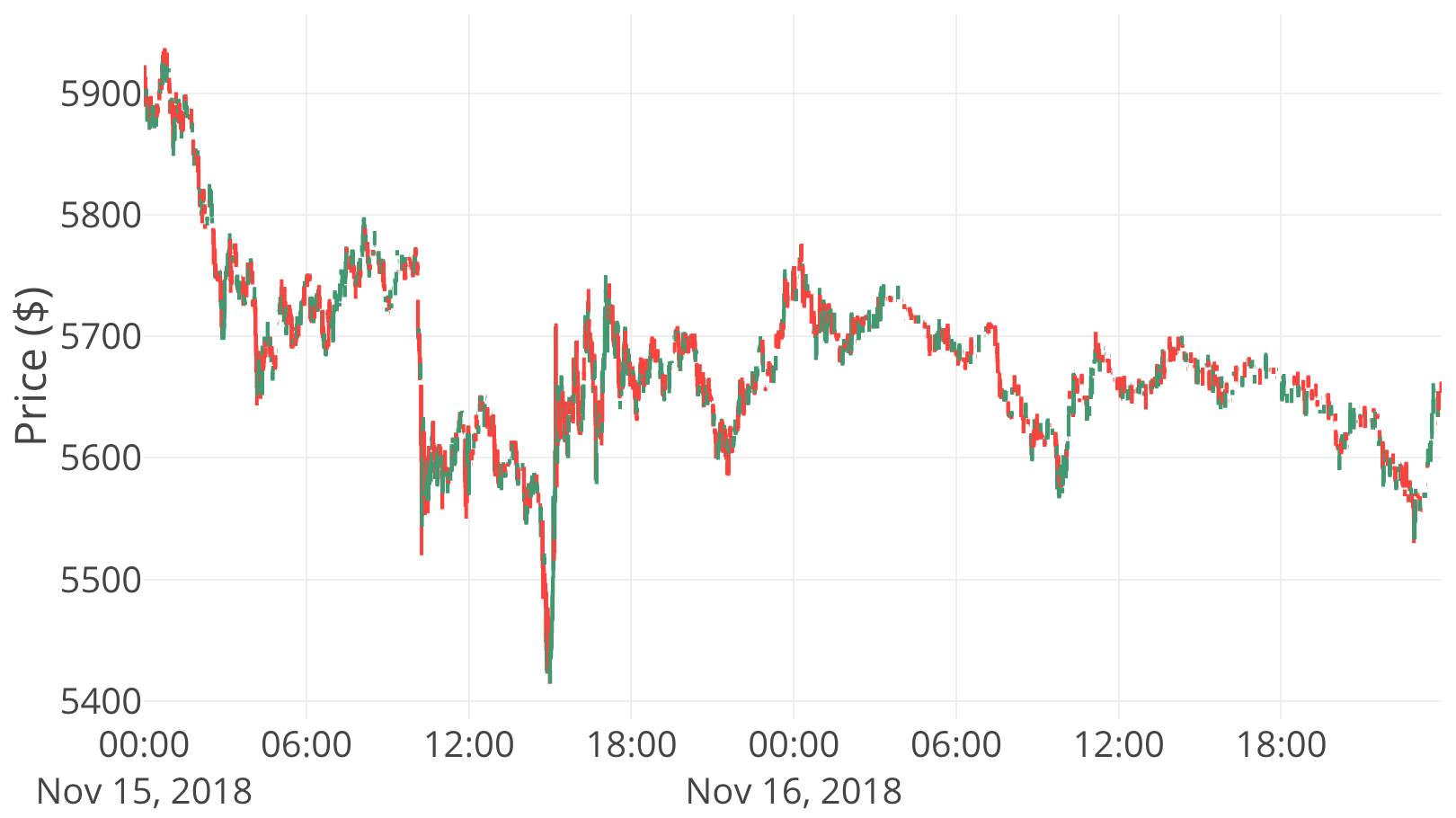}
    \caption{Bitcoin Minute tick data (2018-11-15 00:00 to 2018-11-17 17:06)}
    \label{fig:btc_min_small}
\end{figure}

The Macro-agent is responsible for making a discrete decision to buy, sell, or hold an asset by looking at the data from a macro level. The data used is minute tick data. Although the data is collected over a span of multiple days, the entire data-set is not used for the reason that the agent may under-explore the more recent and relevant states, and over-explore other states. Since most of the data had little volatility, while the more recent data was much more volatile, the agent would not have been able to perform well in volatile states due to exploring the less volatile states more often. Therefore, the recent minute tick data spanning the course of the last few days is used: November 15th, 2018 to November 17th, 2018. Fig. \ref{fig:btc_min_small} shows the portion of the data that was chosen.

\noindent To train the agent, the dataset is split into training and test sets. The training set consists of the time-steps prior to November 16th, 2018. 
\subsection{State}
The state space at a certain time-step $t$ consists of historical price data in the range $t-h$ to $t$, where $h$ denotes how far back in history the agent looks. Furthermore, featurized data of various momentum and reversion indicators are also incorporated into the state space. These indicator features are computed in the following ways:

\vspace{5pt}
\noindent\textit{\underline{Market Indicator Features}}
\vspace{2pt}

One commonly used feature in market trading is the $z$-score indicator, which determines how far (in standard deviations $\sigma$) a data point $x$ is from the mean $\mu$: \begin{align*}z_x = \frac{x-\mu}{\sigma}\tag{5.1}\end{align*} In a way, the $z$-score indicator reveals anomalies in data. It has also been proven to be a useful indicator in determining future trends \cite{c5}. Equation $(5.1)$ can be applied by using an expanding window approach with a window size of $n$ time-steps. Let $p_t$ denote the closing price at time-step $t$, let $\operatorname{SMA}(p_{t-n,t})$ denote simple moving average (SMA) of closing prices over $n$ time-steps prior to $t$, and let $\operatorname{STDDEV}(p_{t-n,t})$ denote the standard deviation of closing prices over $n$ time-steps prior to $t$. Thus, the $z$-score at time-step $t$ for a window size of $n$ is
$$z_{t}^n = \frac{p_{t} - \operatorname{SMA}(p_{t-n,t})}{\operatorname{STDDEV}(p_{t-n,t})}$$
The $z$-score indicator can be extended to volume as well. Thus, the following are the features that are extracted:
\begin{itemize}
    \item \textit{Price Level:} To determine price levels, the $z$-scores are calculated for the prices. This essentially expresses how far each of the prices in the time period are from the average price in the time period.
    \item \textit{Price Change:} To determine price changes, the current price is compared to the average of a window of prices prior to it, i.e. calculated for a window size $n$ at time-step $t$ $$\operatorname{PC}_{t}^n = \frac{p_t}{\operatorname{SMA}(p_{t-n,t})} - 1$$ and take the $z$-score of the result.
    \item \textit{Volume Level:} To determine volume levels, the $z$-scores are calculated for the traded volumes. As with price levels, this expresses how far the volume at each time step is from the volume in the time period.
    \item \textit{Volume Change:} Similar to price change, volume change for a window size $n$ at time-step $t$ is calculated: $$\operatorname{PC}_{t}^n = \frac{v_t}{\operatorname{SMA}(v_{t-n,t})} - 1$$
    and the $z$-score of the result is taken.
    \item \textit{Volatility:} To determine volatility, exponential moving averages (EMA) are used to determine the rate of price change over a span of $n$ days. The reasoning behind using EMA instead of SMA is that EMA gives more weight to the recent time-steps, whereas in SMA all time-steps are weighted equally. Let $p_t$ be the current price, and let $n$ be the window size. The EMA at time-step $t$ is given by $$\operatorname{EMA}_{t}^n = \frac{2p_t}{n+1} + \frac{\sum_{j=t-n}^tp_j}{n}\left(100-\frac{2}{n+1}\right)$$ Thus, the volatility over $m$ days is given by $$\operatorname{Volatility}_{t}^m = \frac{\operatorname{EMA}_{t,n} - \operatorname{EMA}_{t-m,n}}{\operatorname{EMA}_{t-m,n}}$$
\end{itemize}

Along with these market indicator features and the price data, there is an additional state parameter which plays an essential role in determining rewards. Each time the agent chooses to buy, the price at which the agent buys the asset (the current open price) is stored into a current assets list. This list of prices is stored as part of the state as well. Each time the agent chooses to sell, all of the bought assets are exhausted (sold). Thus, the historical prices, market indicator features, and current assets list form the state.
\subsection{Action} There are three actions that the agent can choose:
\begin{enumerate}
    \item \textit{Buy:} If the decision to buy is chosen, then the agent is choosing to buy $1$ bitcoin at the current opening price. As mentioned previously, when the agent chooses to buy, the price gets appended to the current assets list. Note that the decision to buy $1$ bitcoin is purely a design choice. The proposed framework can be applied to any quantity of bitcoins.
    \item \textit{Sell:} If the decision to sell is chosen, then the agent is choosing to sell all of its accrued assets at the current opening price. 
    \item \textit{Hold:} If the decision to hold is chosen, then the agent does not do anything.
\end{enumerate}
\subsection{Reward}
Upon taking an action, the agent receives a reward from the environment. Algorithm 1 depicts how rewards are handled. 

\begin{algorithm}
    \uIf{action $=$ hold} {
        $reward\leftarrow 0$
    } 
    
    \uElseIf{action $=$ buy}{
        \textbf{Append} current open price to current assets list
        
        $reward\leftarrow 0$
    }
    
    \uElseIf {action $=$ sell}{
        \eIf{there are no assets to sell}{
            $reward\leftarrow -1$
        } {
            $reward\leftarrow $ sell off all assets from current assets list and determine profit based on current open price
        }
        
    }
    
    \textbf{Clip Rewards}
    
    \caption{Macro-agent Reward Function}
\end{algorithm}

\noindent Note that the rewards are clipped to $\{-1,0,1\}$ based on negative, zero, or positive rewards, respectively. This is done in order to have the agent deal with different types of market environments, e.g. high/low volatility, as well as reduce the impact of anomalous market shifts. Clipping rewards has proven to be a useful method when dealing with different scales of rewards when in various Atari games \cite{c6}.

\subsection{Deep $Q$-Network Architecture \& Training}
\begin{figure}[!ht]
    \centering
    \includegraphics[scale=0.19]{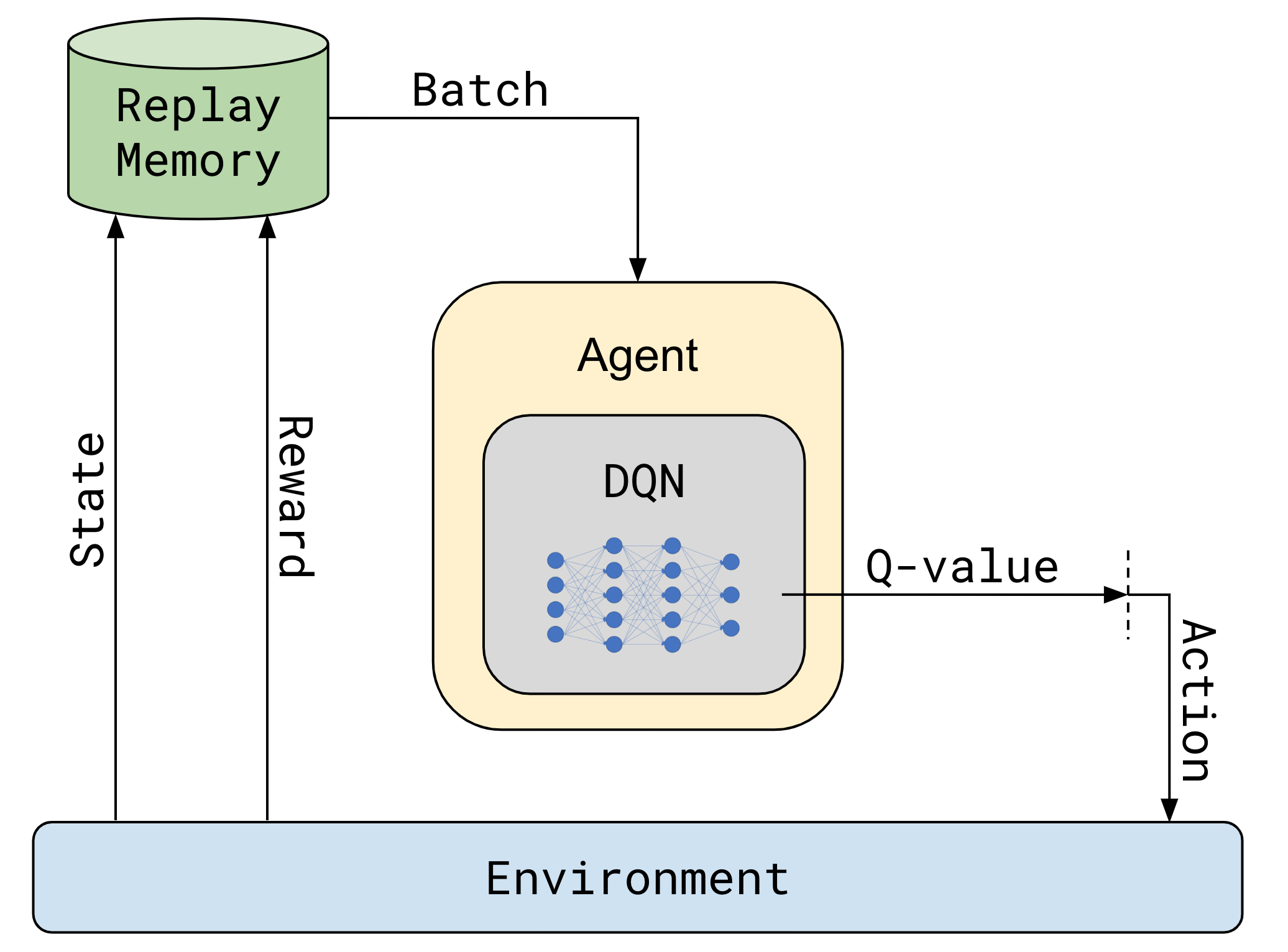}
    \caption{Macro-agent Training Framework}
    \label{fig:macro_arch}
\end{figure}

To parameterize the value function, a deep $Q$-network is used. A simple multilayer perceptron with two hidden layers (with ReLu activation functions) is chosen. Furthermore, the Adam optimizer is used for training. The $Q$-network takes in as inputs the price data, market indicator data, and the current assets list, and outputs $3$ $Q$-values which are then used to determine the optimal action. In Fig. \ref{fig:macro_arch} a depiction of the training framework for the macro-agent is provided. 

The details of how training happens within this framework are outlined in Algorithm 2. Note that for each epoch the agent is trained on the entire training set. 

\begin{algorithm}
    \textbf{Initialize} replay memory $M$
    
    \textbf{Initialize} $\epsilon$, and discount factor $\gamma$
    
    \textbf{Initialize} $Q$ network with random weights
    
    \For{$\text{epoch}=1$ to $E$}{
        $s_t\leftarrow$ \textbf{Reset} environment (environment outputs initial state)
        
        \While{not done} {
            $a_t\leftarrow$ \textbf{Select} random action with probability $\epsilon$ or choose action $\max_a Q(s,a;\theta)$
            
            $s_{t+1},r_t,done\leftarrow$ \textbf{Act} based on $a$ (environment outputs new state, reward, and done)
            
            \uIf{replay memory $M$ is full} {
                \textbf{Remove} the first element from $M$
            }
            
            \textbf{Append} $(s_t,a_t,r_t,s_{t+1},done)$ to replay memory $M$

            $b\leftarrow$ \textbf{Sample} mini-batch from replay memory $M$ at random
            
            $q\leftarrow$ \textbf{Initialize} empty array of size $|b|$
            
            \ForEach{$(s_i,a_i,r_i,s_{i+1})\in b$} {
                $q_i = \begin{cases}r_i + \gamma\cdot\max_a Q(s_{i+1},a_i;\theta) & \text{not } done\\r_i & done\end{cases}$
                
            }
            
            \textbf{Apply} gradient descent with loss $\mathbb{E}_{s,a,r,s'}\left[(q_i - Q(s_i, a_i;\theta))^2\right]$

            \textbf{Decay} $\epsilon$
        }
    }
    \caption{Macro-agent Deep $Q$-Learning Training}
\end{algorithm}

\subsection{Results}
It is interesting to see how the macro-agent would perform on its own. Although the closing and opening prices, and the prices used to calculate profits for the macro-agent are usually not indicative of the true prices where trades may happen, they can hypothetically be assumed so to evaluate the performance of the agent. For these results, the agent was trained on the training set and tested on the test set, as defined previously, for $500$ epochs.

To understand the potential of the macro-agent, its performance is compared to two common investment strategies:
\begin{itemize}
    \item \textit{Buy and Hold investing:} Buy and Hold investing is a naive long-term investment strategy where an investor buys an asset at the start of a time period and holds the asset in the hopes that it will accrue value over time. From the looks of Fig. \ref{fig:btc_min_small} it is not expected that this strategy will yield any positive profit as the general trend in the test set is downwards. In this strategy, $10$ Bitcoins are chosen to be bought and held over the time period. In order to simulate the Buy and Hold investing strategy, at each time-step the profit is plotted assuming that the investor decided to sell at each time-step, i.e. if $p_{0}$ is the price the asset was initially bought at, at each time-step the difference $p_t - p_0$ is plotted.
    \item \textit{Momentum investing:} Momentum investing is where an investor chooses to buy or sell an asset at a certain time-step given the performance of the asset in the last $n$ steps. To implement this, a simple moving average of the prices with a window size of $n = 20$ is computed, i.e. $20$ minutes prior. At each time-step, if the current open price is less than the average price in the minutes prior, $1$ Bitcoin is bought. If the current open price is greater than the average price in the minutes prior, all Bitcoins bought thus far are sold. In the case they are equal, all assets are held.
\end{itemize}

\begin{figure}[!ht]
    \centering
    \includegraphics[scale=0.30]{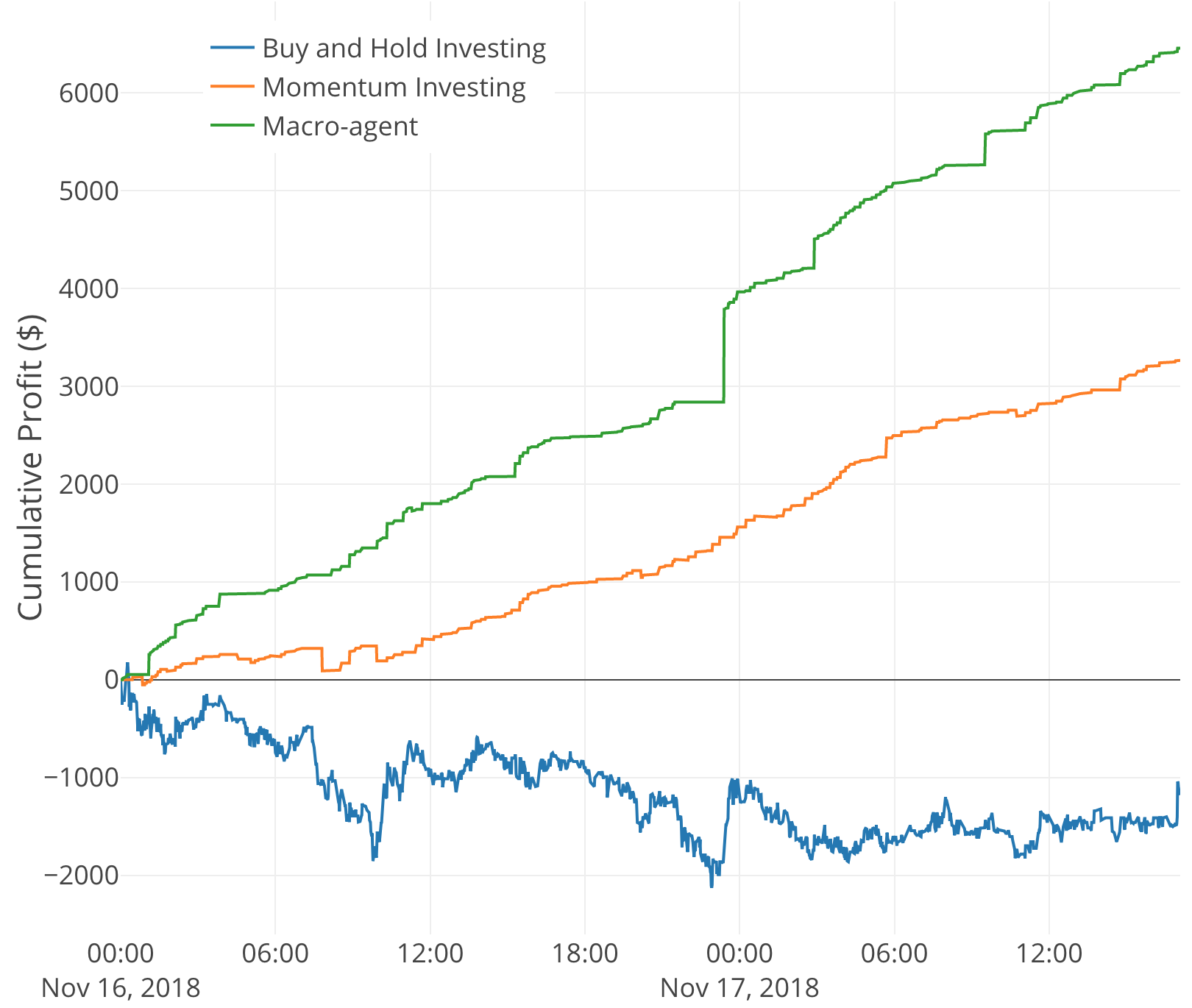}
    \caption{Performances of Various Investment Strategies Realized-PNL Graph}
    \label{fig:macro_perf}
\end{figure}

These investment strategies serve as benchmarks and are compared with the macro-agent. The macro-agent is similar to momentum investing with the only difference being that the policy of when to buy, sell, or hold is different. Similar to momentum investing, the macro-agent buys one bitcoin on a decision to buy and exhausts all assets on a decision to sell. Fig. \ref{fig:macro_perf} depicts the performance comparisons between the different strategies.

Fig. \ref{fig:macro_perf} compares the strategies in terms of profit. The accumulated profit is plotted for the momentum investing strategy and the macro-agent. The macro-agent's accumulated profit growth is not very volatile, indicating that the macro-agent strategy is stable as well.

\section{Details of the Mirco-agent}
In the previous section, the results indicated that the macro-agent performed well compared to two known investment strategies. However, one issue with the macro-agent is that it assumes that the trade occurs at the opening prices. In real market environments a trade happening at the opening price is most often not the case. Therefore, there lies this uncertainty in the framework thus far of what price the order should be set at. This motivates the purpose for the micro-agent. Many of the ideas used here are derived from \cite{c3} and \cite{c4}.

\subsection{Introduction}
\begin{figure}[!ht]
    \centering
    \includegraphics[scale=0.16]{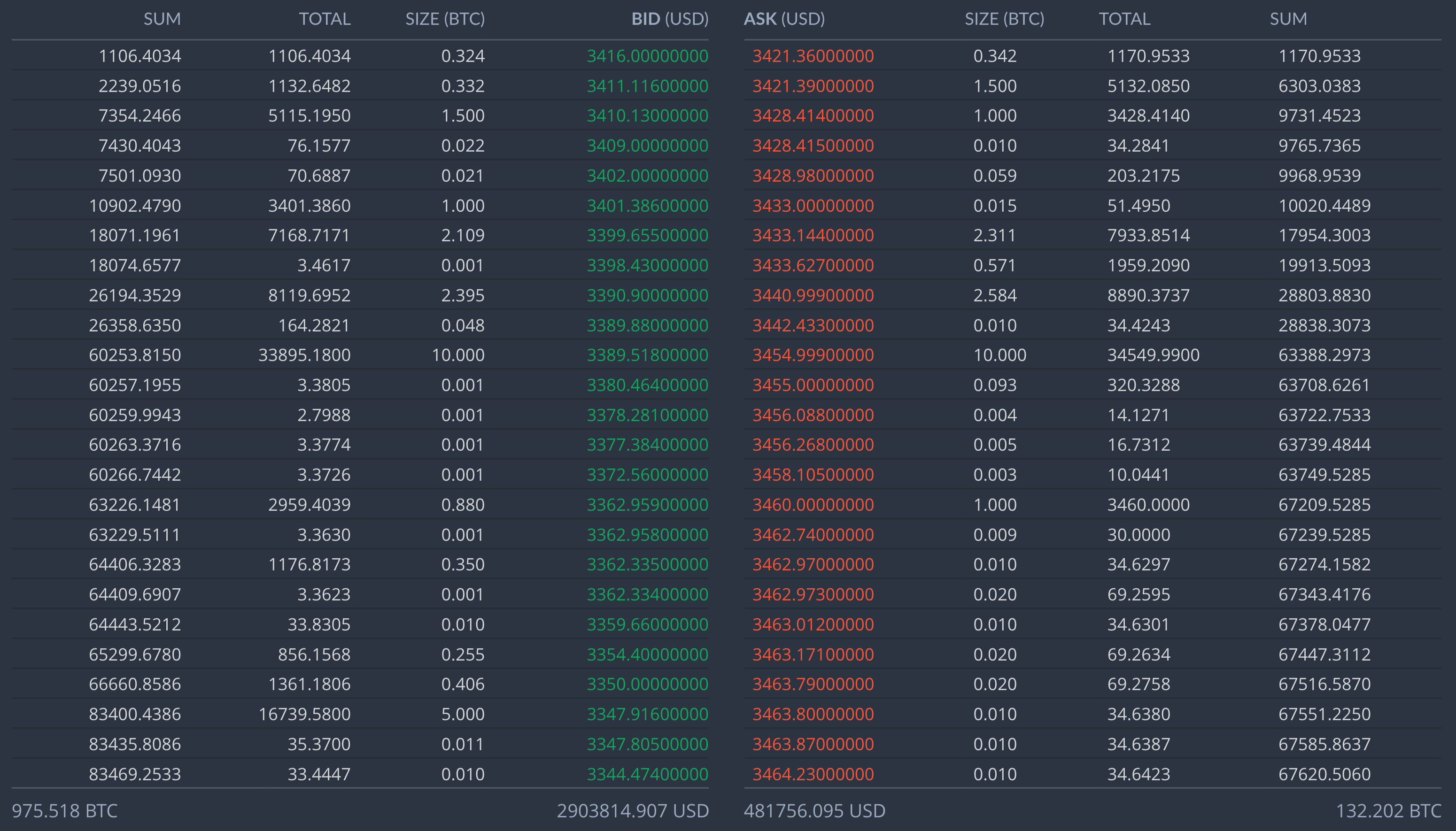}
    \caption{Example of an Order Book State}
    \label{fig:order book}
\end{figure}

In financial markets, there is an order book, a data structure that holds all bid and ask price/volume values. Fig. \ref{fig:order book} depicts a snapshot of an order book state.

\noindent The order book is separated into two parts: bids and asks. The bid section contains all the prices and quantities buyers are willing to buy at. The ask section contains all the prices and quantities sellers are willing to sell at.  There are two main ways that a trader can place an order in the order book:
\begin{enumerate}
    \item \textit{Market Order:} When a trader places a market order, they are placing an order at the market price, i.e. the best price of the opposing side of the order book. Most often, this order is immediately filled. However, depending on the quantity the order was placed for, the order might not be executed at the same prices, as it depends on how much quantity is available at market price. Another disadvantage is that market orders usually come with additional trading fees. Depending on the state, this may even yield negative profit.
    \item \textit{Limit Order:} When a trader places a limit order, they are essentially demanding the order price (or a better price). A limit order guarantees the trader the order price, however, it is possible for the order to never be filled. When a trader places a limit order, they stand behind all the traders that have placed an order at that price. An advantage to placing limit orders is that since limit orders provide liquidity to other market participants, exchanges incentivize limit orders by removing additional trading fees.
\end{enumerate}
Once a trader places an order, a match engine attempts to match the trader's order to orders on the opposing order book side. Only when there is a matching order does a trade happen.

The agent should be able to optimally place orders in the order book. As noted previously, the agent has two order options: market orders or limit orders. However, there is a trade-off between placing market orders and placing limit orders. Therefore, the goal of the agent is to optimize on this trade-off: is there an action better than simply placing a market order at the beginning of the time horizon?

This is a challenging optimization problem since there are many unknown variables here:
\begin{itemize}
    \item \textit{Other Market Participants:} This agent is not the only one trading on the market. There are many other market participants that the agent interacts with.
    \item \textit{Adversarialness:} Among these market participants, there are also algorithmic traders, some of which might be playing adversarially, e.g. Bitcoin whales.
\end{itemize}

The agent must also optimize over these unknowns as well. In order to do that, the match engine needs to accurately simulate the market. Since this problem is posed as an RL problem, such a simulator is essential, since the agent requires reward feedback. An open-source matching engine \cite{c7} is used to help simulate the exchange match engine. However, since the data is limited to the time periods it was collected for, the match engine simulator is only indicative to that time period. Therefore, there is still a disadvantage to using an artificial match engine. 

\subsection{Environment}
In reinforcement learning problems, the environment is essential since it provides the agent with subsequent states, as well as reward feedback. The environment for this problem is vastly different than the macro-agent's environment. The environment takes in as input the order side, i.e. buy or sell from the macro-agent.

The two major components of the environment are the matching engine and the order book. The order book plays a key role in the environment. Bid, ask, and trade data are collected from the WebSocket. Additionally, at the start of data collection, a snapshot of the current order book state for the first $20$ levels on each side are taken. Based on this snapshot, and the bid and ask data alone, the entire historical order book is created.

The data is again split into a training and test set similar to the macro-agent. During training, the environment starts at a random time-step within the training set (at the start of a minute). The environment lasts until the end of the minute. Within the minute, the agent chooses to place an order which the matching engine attempts to match. If for some reason the order has not been matched until the last time-step, the environment forces a market order for all remaining assets. The environment provides reward feedback only when a trade has occurred, or when the time horizon has been exhausted. This single interaction is defined as one epoch.

\subsection{State}
As part of the state, the concept of private and market variables is used:
\begin{itemize}
    \item \textit{Private Variables:} Private variables contain two features: quantity remaining, and time remaining.
    \item \textit{Market Variables:} Market variables contain order book states and trade data from the past $30$ time-steps. An order book state consists of the bid and ask prices and quantities up to $20$ levels. Trade data consists of the price and quantity, and order side at which the last trade occurred.
\end{itemize}

\subsection{Action}
The action the agent decides is the price at which to place a limit order. More specifically, the agent chooses an integer action $a\in[-50,50]$, such that the price, $p_t$, at time-step $t$ is $$p_t = p_{m_t} + 0.10a$$ where $p_{m_t}$ is the market price. Thus, there are a total of $101$ possible discrete actions the agent can choose from.

\subsection{Reward}
Recall that the goal of the micro-agent is to optimally place limit orders. Therefore, the agent is rewarded accordingly. The reward function is the difference between what the agent was able to sell at and the market price before the order was placed. The Volume Weighted Average Price (VWAP) is used to determine the aggregated price the agent was able to trade at. Thus, the reward function is $$R_t = p_{m_t} - \frac{1}{\sum_i v_i}\sum_{p\in P} v_p\cdot p$$ where $p_{m_t}$ is the market price, $\sum_i v_i$ is the total volume of assets, and $P$ is all the prices the agent ordered at.

\subsection{Deep $Q$-Network Architecture \& Training}
For this agent, a similar training framework to the macro-agent is used. However, a different network architecture is used.

\vspace{5pt}
\noindent\textit{\underline{Dueling Deep $Q$-Network}}
\vspace{2pt}

For many states, the actions $a$ on the edge of the order book may have little value to the states. Therefore, for some states, it may be unnecessary to estimate the action values for all actions. Dueling Deep $Q$-Networks (DDQNs) provide a solution to this. First applied to Atari games, DDQNs not only help to speed up training, but also yield more reliable $Q$-values \cite{c8}.

\begin{figure}[!ht]
    \centering
    \includegraphics[scale=0.17]{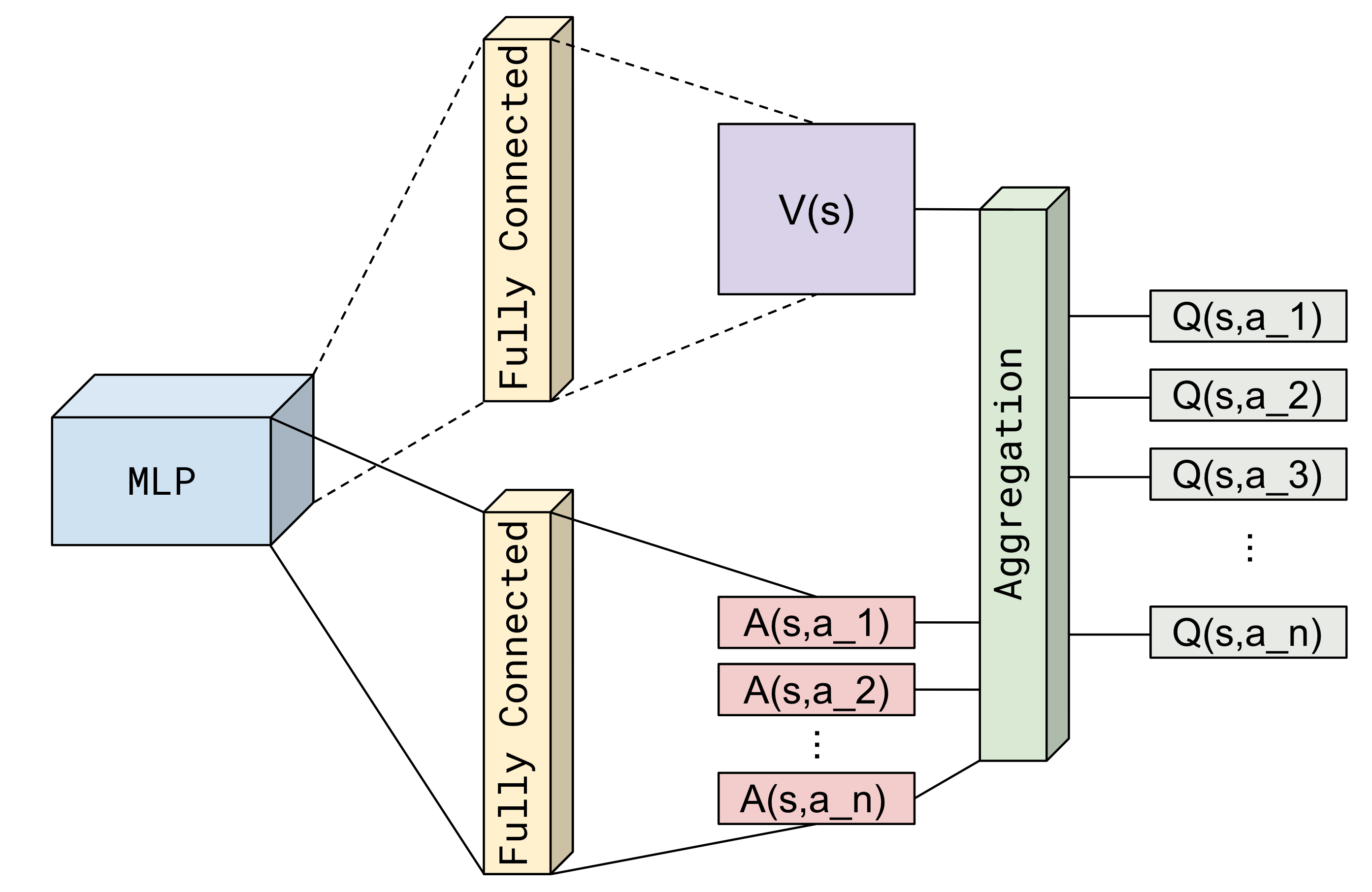}
    \caption{Dueling Deep $Q$-Network Framework}
    \label{fig:nn}
\end{figure}

Recall that a $Q$-value can be decomposed into two parts: $V(s)$, the value of being in a particular state $s$, and $A(s,a)$, the advantage of taking action $a$ in state $s$, i.e. the $Q$-value is decomposed as $$Q(s,a) = V(s) + A(s,a)$$ DDQNs decouple the prediction into two network streams (Fig. \ref{fig:nn}): value stream, which estimates $V(s)$, and advantage stream, which estimates $A(s,a)$. This decoupling allows the agent to learn which states might not be valuable enough to explore every action. After this decoupling, there is an aggregation layer which aggregates the two streams by the following:
\begin{align*}Q&(s,a;\theta,\alpha,\beta) \\&= \hat{V}(s;\theta,\beta) + \left(\hat{A}(s,a;\theta,\alpha) - \frac{1}{|A|}\sum_{a'}\hat{A}(s,a';\theta,\alpha)\right)\end{align*}
where $\theta$ are the weights for the MLP, $\alpha$ are the weights for the advantage stream, and $\beta$ are the weight for the value stream. 

\begin{algorithm}
    \textbf{Initialize} replay memory $M$
    
    \textbf{Initialize} $\epsilon$, discount factor $\gamma$
    
    \textbf{Initialize} $Q$ network with random weights

     \For{$\text{epoch}=1$ to $E$}{
        $s_t\leftarrow$ \textbf{Reset} environment (environment outputs random state and gets inputs from macro-agent)
        
        \While{$t \leqslant 60\text{ seconds}$} {
            $a_t\leftarrow$ \textbf{Select} random action with probability $\epsilon$ or choose action $\max_a Q(s,a;\theta)$
            
            $s_{t+1},r_t,done\leftarrow$ \textbf{Act} based on $a$ (environment outputs new state, reward, and done, based on match engine output)
            
            \uIf{replay memory $M$ is full} {
                \textbf{Remove} the first element from $M$
            }
            
            \textbf{Append} $(s_t,a_t,r_t,s_{t+1},done)$ to replay memory $M$

            $b\leftarrow$ \textbf{Sample} mini-batch from replay memory $M$ at random
            
            $q\leftarrow$ \textbf{Initialize} empty array of size $|b|$
            
            \ForEach{$(s_i,a_i,r_i,s_{i+1})\in b$} {
                $q_i = \begin{cases}r_i + \gamma\cdot\max_a Q(s_{i+1},a_i;\theta) & \text{not } done\\r_i & done\end{cases}$
                
            }
            
            \textbf{Apply} gradient descent with loss $\mathbb{E}_{s,a,r,s'}\left[(q_i - Q(s_i, a_i;\theta))^2\right]$

            \textbf{Decay} $\epsilon$
        }
    }
   
    \caption{Micro-agent Deep $Q$-Learning Training}
\end{algorithm}

The concept of dueling deep $Q$-networks is applied to the micro-agent. Besides the dueling concept, the rest of the training framework remains the same (see Fig. \ref{fig:marl}). The modifications to Algorithm 2 are detailed in Algorithm 3.

\subsection{Results}
In this section, the performance of the micro-agent strategy is evaluated. The basic goal behind this agent is to optimally place limit orders to buy or sell an asset within the allotted time horizon. Here, the agent is tested through two scenarios, each evaluating its performance on ideal and un-ideal states.
\subsubsection{Market Trending Downwards} In this scenario a time-step is chosen within the test set where the market is generally trending downwards. Here, the ideal case is to buy an asset, since the price is generally decreasing.

\begin{figure}[!ht]
    \centering
    \includegraphics[scale=0.30]{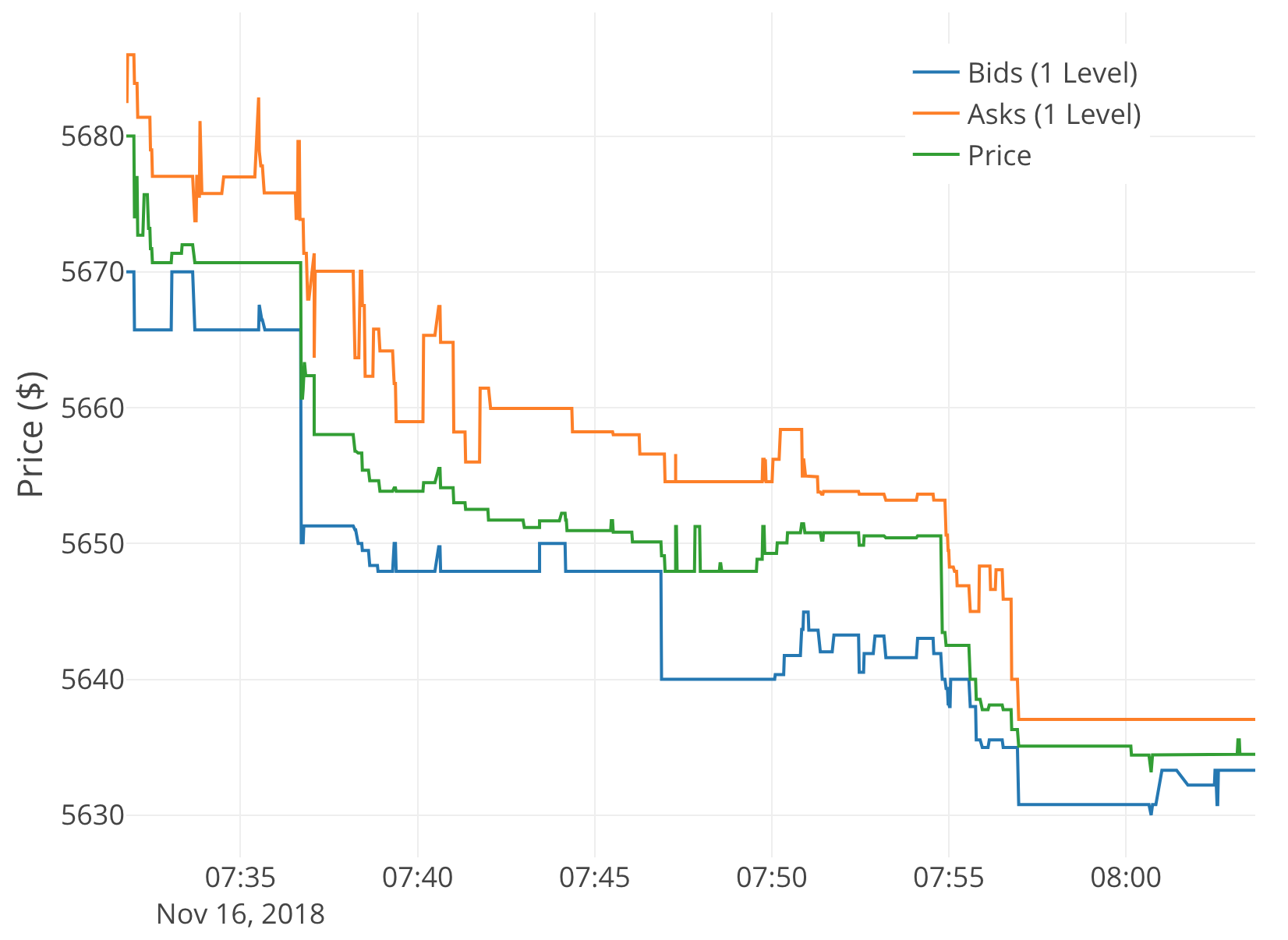}
    \caption{Order Book Visualization for Downward Trend}
    \label{fig:down}
\end{figure}

Fig. \ref{fig:down} depicts the snapshot chosen to test the agent against a downward trend. More specifically, the agent is tested on the \texttt{07:36} to \texttt{07:37} time horizon.

\vspace{5pt}
\noindent\textit{\underline{Buying an Asset}}
\vspace{2pt}

When buying an asset on an upward trend, the agent decided to take an action of $-11$. This meant setting a limit order to buy at $\$1.10$ lower than market price. This resulted in immediate execution. This was the right choice to make, since the market was trending downward.

\vspace{5pt}
\noindent\textit{\underline{Selling an Asset}}
\vspace{2pt}

When selling an asset on a downward trend, the agent decided to take an action of $29$. This amounted to a limit order price of $\$2.90$ greater than the market price. Although this may not seem like the optimal decision, given that the agent was able to recognize the price drop, it decided to immediately sell off its asset. At such a price it is almost guaranteed for a trade to occur, as was the case here. The agent was able to sell off $1$ bitcoin in the first step (before the market dropped).

\vspace{8pt}

\subsubsection{Market Trending Upwards} In this scenario a time-step within the test set is chosen where the market is generally trending upwards. Here, selling an asset is the ideal case, since the price is generally increasing.

\begin{figure}[!ht]
    \centering
    \includegraphics[scale=0.30]{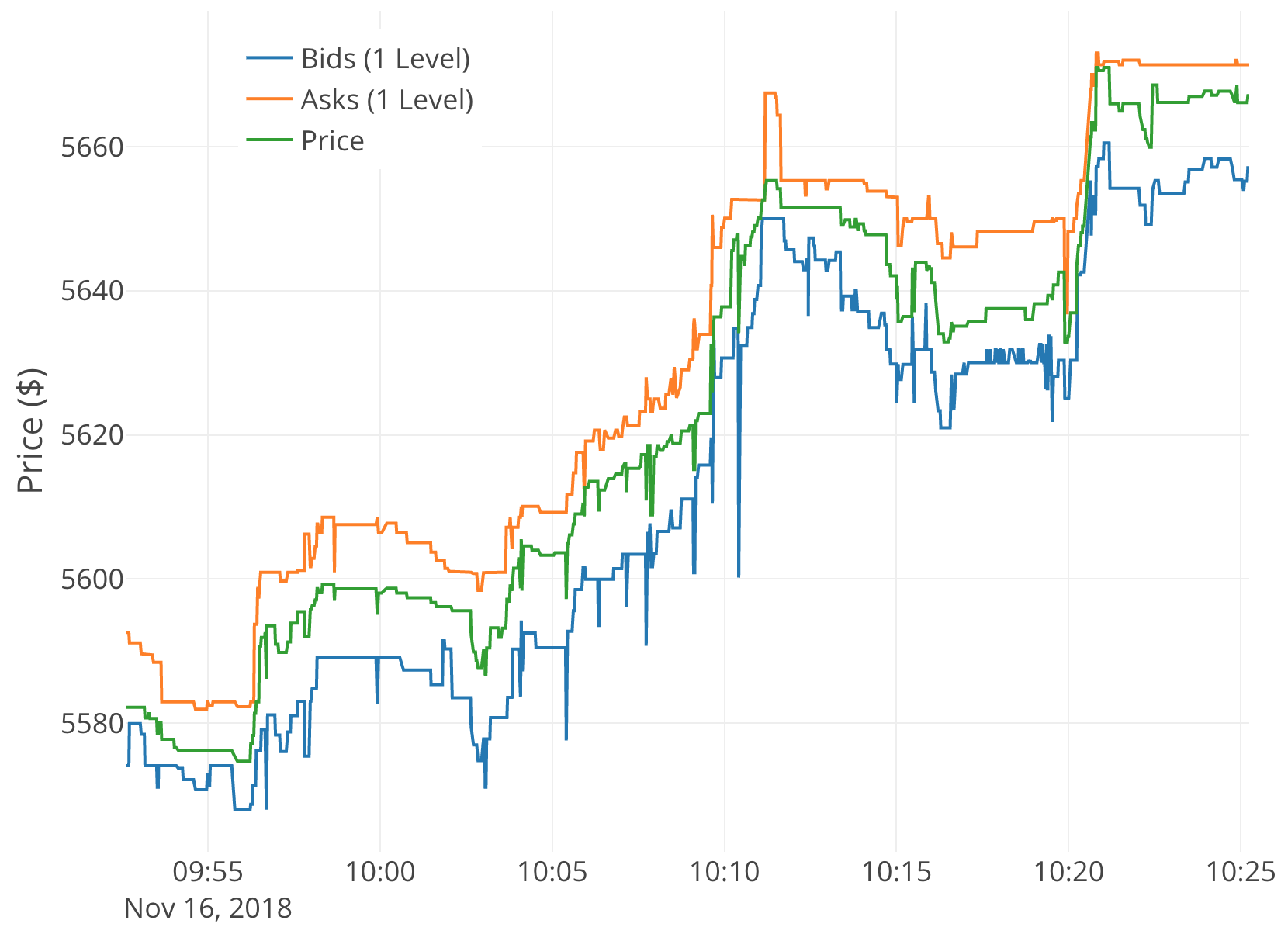}
    \caption{Order Book Visualization for Upward Trend}
    \label{fig:up}
\end{figure}

Fig. \ref{fig:up} depicts the snapshot chosen to test the agent on an upward trend. More specifically, agent was tested on the \texttt{10:09} to \texttt{10:10} time horizon.

\begin{figure*}[!ht]
    \centering
    \includegraphics[scale=0.25]{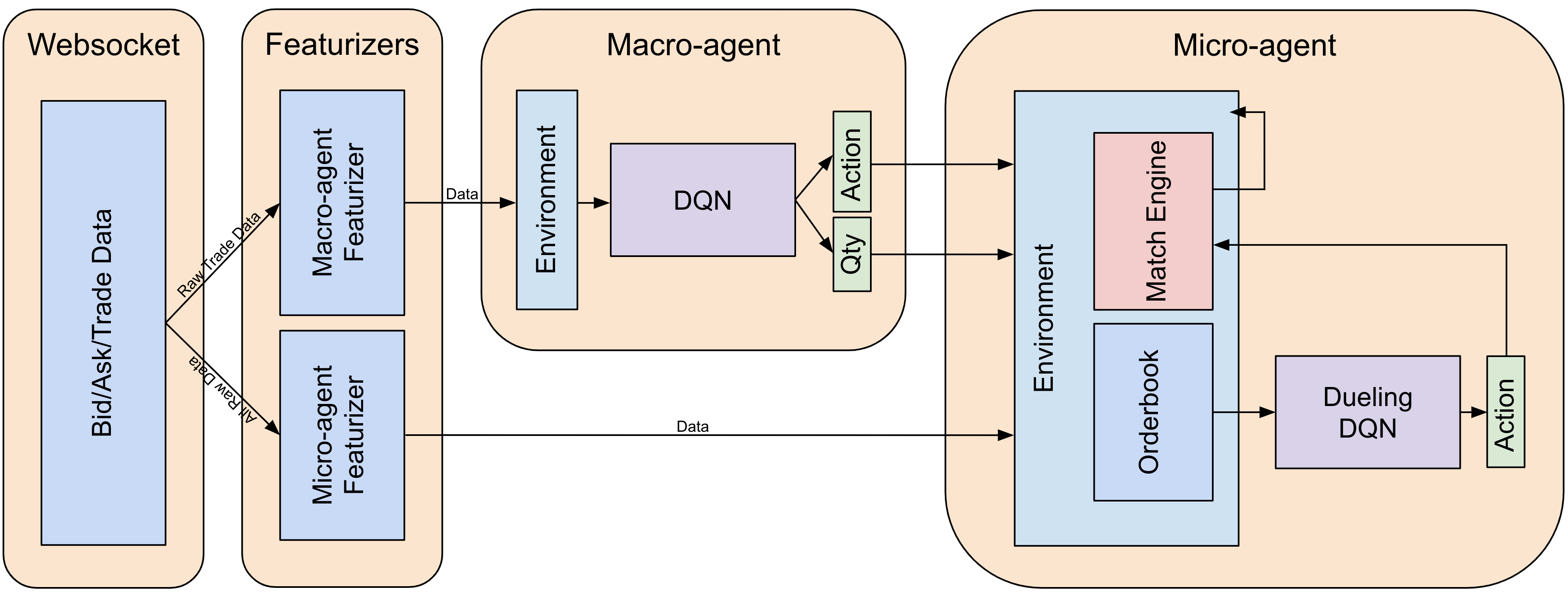}
    \caption{Combined Multi-agent End-to-End Pipeline}
    \label{fig:combined}
\end{figure*}

\vspace{5pt}
\noindent\textit{\underline{Buying an Asset}}
\vspace{2pt}

When buying an asset on an upward trend, the agent decided to take an action of $18$. This meant setting a limit order to buy at $\$1.80$ higher than market price. This resulted in immediate execution. This was the right choice to make, since the market was trending upwards.

\vspace{5pt}
\noindent\textit{\underline{Selling an Asset}}
\vspace{2pt}

When selling on an upward trend, the agent decided to take an action of $-21$. This meant setting a limit order to sell at $\$2.10$ lower than market price. This was a risky decision, since the order may not have been matched. In fact, it took two steps for the agent to sell $1$ bitcoin, i.e. it had to place $2$ limit orders (each partially executed). In this case a negative action was the right choice to make since the price was increasing.

Based on these special cases, the micro-agent looks to be able to make optimal decisions. When the price is expected to drop, the agent makes the optimal decision to sell off all assets immediately (or limit buy assets at a lower price than the market on the bid side). On the other hand, when the price is expected to grow, the agent makes the optimal decision to buy assets immediately (or sell assets at a lower price than the market on the ask side). 

\section{Combining the Macro and Micro agents}

In Sections 5 and 6, an in-depth overview of the frameworks for both the Macro and the Micro agents was given. Now, the two agents are combined to reflect the multi-agent framework defined in Fig. \ref{fig:marl}. The end-to-end pipeline contains $4$ components, depicted in Fig. \ref{fig:combined}, which is briefly detailed as follows:
\begin{enumerate}
\item \textit{WebSocket:} Raw trade, bid, and ask data arrives through a WebSocket that is connected to the Bittrex exchange. As soon as data arrives, it is sent to the featurizers.
\item \textit{Featurizers:} There are two featurizers, one for each agent. The Macro-agent featurizer computes the market indicator features and the minute data based on the raw trade data collected from the WebSocket. The Micro-agent featurizer determines the new state of the order book based on the raw data.
\item \textit{Macro-agent:} As detailed in Section 5, the macro-agent determines the action to buy, sell, or hold the asset. The macro-agent keeps a running count of how many assets it currently has. If the decision to sell is chosen, then this count (total number of assets) is sent. If the decision to buy is chosen, then a quantity of $1$ is sent. 
\item \textit{Micro-agent:} As detailed in Section 6, the micro-agent determines the action of where to place the order in the limit book. The side of the order book and the quantity is determined by the data provided by the macro-agent. Once the micro-agent has decided an action, this action gets fed into the match engine which attempts to match the order. This decision is relayed back to the environment for the agent to decide to cancel the order and create a new order. If the order is matched, i.e. turns into a trade, then the order book is updated. Note that in practice the match engine is the actual exchange.
\end{enumerate}
\subsection{Final Results}
The end-to-end pipeline is evaluated using the test set defined in Section 5. Similar to the evaluation of the macro-agent, the multi-agent strategy is compared to the Buy and Hold investing and Momentum investing strategies.

\begin{figure}[!ht]
    \centering
    \includegraphics[scale=0.30]{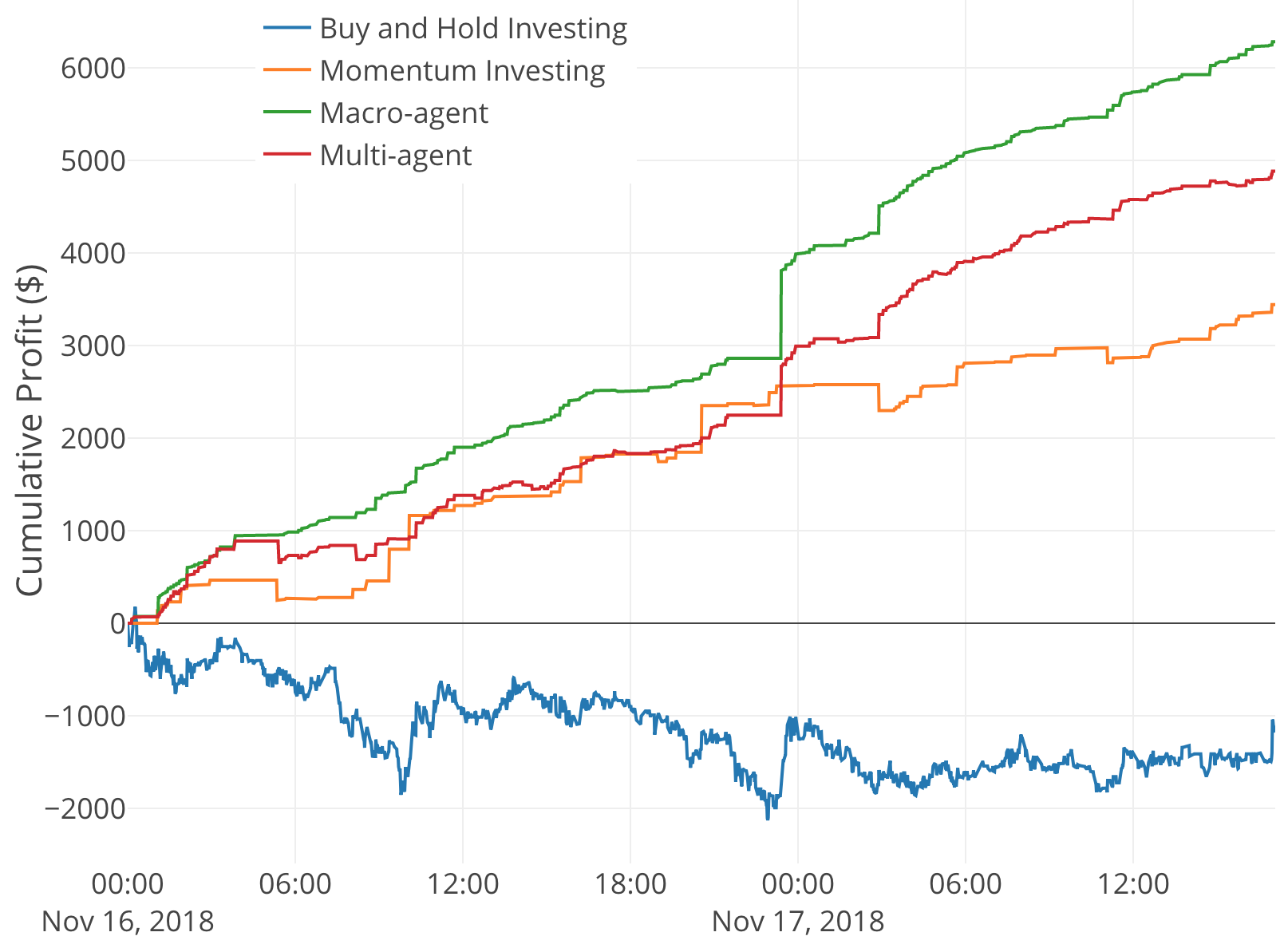}
    \caption{Performances of Various Investment Strategies PNL-Realized Graph (including Multi-agent)}
    \label{fig:finalresult}
\end{figure}

In Fig. \ref{fig:finalresult} the performance comparisons between the Buy and Hold investment, Momentum investment, Macro-agent, and Multi-agent (Macro-agent and Micro-agent) strategies are given. It is interesting to see that the multi-agent approach under-performs in comparison to the macro-agent. Previously, in the evaluation of the micro-agent, it was noticed that often times the agent would decide that the right decision was to place an order towards the opposing side of the order book, i.e. a price slightly worse than market price. Although this resulted in immediate execution, this resulted in a drop in cumulative profit. 

In terms of total orders placed by the micro-agent in the multi-agent setting, $91\%$ of all orders placed were limit orders. Thus, the micro-agent was able to optimize placing limit orders. Note that in the other strategies hypothetical prices were assumed for which market orders would have been placed for. This would also result in additional exchange fees. 

Overall, the proposed multi-agent framework is able to optimize well. Additionally, the multi-agent approach is also a stable strategy since the cumulative profit growth is not volatile.

\section{Improvements \& Future Work}
Although considerable progress was made in creating an end-to-end framework for optimizing market making, there are still some areas of improvement:
\begin{itemize}
\item \textit{Risk:} One assumption made throughout this study was that the agent was allowed to buy an unlimited number of assets. There is high risk associated with this decision, since the agent is in control of how many assets to buy. When testing the multi-agent strategy, it was found that there was a case when the agent was selling $84$ bitcoins with relatively low profit. The risk associated with this is also high. The next step would be to optimize on this by adding further constraints that limit the number of buy actions the macro-agent takes. However, this requires the agent to better learn when to hold an asset. This gives rise to the next concern.
\item \textit{Reward Engineering:} Reward engineering is an important problem in reinforcement learning. In the case of the macro-agent, it was found that the decision to hold an asset was very sparse. This may have resulted from the agent receiving a reward of $0$ for the decision to hold. Perhaps setting the constraints as previously mentioned along with a reward function that is able to provide better reward feedback for holds, may help the agent learn to hold assets for a longer time. Furthermore, incorporating exchange fees in the reward function will lead to an agent that can successfully interact with the exchange environment.
\item \textit{Market Simulator:} Although the simulator used in this framework did provide a decent reflection of the true market environment, it still lacked many variables. Specifically the effect of other traders. The assumption that the micro-agent was the only trader interacting with the order book was also made. One option here is to create a generative model that can create synthetic orders that simulate various traders.
\item \textit{Corrupted Data:} Another concern is corrupted data. On multiple instances, it was found that data would arrive out of order, or only partial (missing) data was received due to network issues. In practice, it is critical that the states match the actual exchange. Therefore, a solution to this is needed before this framework can be deployed on an actual exchange.
\end{itemize}

\section{Conclusion}
A multi-agent reinforcement learning framework was provided to solve the problem of optimizing market making. The results showed that the policy these agents were able to learn led to a stable trading strategy which resulted in a low-volatile linear growth in profit. Applying reinforcement learning to such a problem also shows that it has the ability to perform well in complex environments, and is a viable tool that can be used for market making.

\end{document}